\documentclass{emulateapj}

\slugcomment{}

\shorttitle{Do all Flares have White Light Emission?}
\shortauthors{D.B. Jess et al.}

\begin{document}

\title{Do all Flares have White Light Emission?}

\author{D. B. Jess}
\affil{Astrophysics Research Centre, School of Mathematics and Physics, Queen's University, Belfast, BT7~1NN, 
Northern Ireland, U.K.}
\affil{}
\affil{NASA Goddard Space Flight Center, Solar Physics Laboratory, Code 671, Greenbelt, MD 20771, USA}
\email{djess01@qub.ac.uk}

\and

\author{M. Mathioudakis, P. J. Crockett and F. P. Keenan}
\affil{Astrophysics Research Centre, School of Mathematics and Physics, Queen's University, Belfast, BT7~1NN, 
Northern Ireland, U.K.}

\begin{abstract}

High-cadence, multiwavelength optical observations of a solar active region (NOAA~10969), obtained with the 
Swedish Solar Telescope, are presented. Difference imaging of white light continuum data reveals a 
white light brightening, 2~min in duration, linked to a co-temporal and co-spatial C2.0 flare event. The 
flare kernel observed in the white light images has a diameter of 300~km, thus rendering it below the 
resolution limit of most space-based telescopes. Continuum emission is present only during the impulsive stage 
of the flare, with the effects of chromospheric emission subsequently delayed by $\approx~2$~min. 
The localized flare emission peaks at 300\% above the quiescent flux. This large, yet tightly confined, increase in emission is 
only resolvable due to the high spatial resolution of the Swedish Solar Telescope. An investigation of the line-of-sight 
magnetic field derived from simultaneous {\sc{mdi}} data shows that the continuum brightening is located very close to a magnetic 
polarity inversion line. Additionally, an H$\alpha$ flare ribbon is directed 
along a region of rapid magnetic energy change, with the footpoints of the ribbon remaining co-spatial with 
the observed white light brightening throughout the duration of the flare. The observed flare parameters are 
compared with current observations and theoretical models for M- and X-class events and we determine the 
observed white-light emission is caused by radiative back warming. We suggest that the creation of white-light emission 
is a common feature of all solar flares.

\end{abstract}


\section{Introduction}
\label{intro}

Solar white-light flares (WLFs) are events which are visible in the optical continuum and normally contain 
very small flare kernels, often less than $3\arcsec$ in diameter \citep[]{Nei89}. Traditionally, 
WLFs have been associated with more energetic solar flares, where the EUV or soft X-ray luminosity exceeds 
a relatively large threshold \citep[]{Nei83}. Large X-class WLFs have been observed with ground- and space-based 
telescopes in recent years \citep[]{Xu06,Iso07}, and studies of WLF energetics have indicated that the 
energy required to power the WL emission is similar to the total energy carried by the electron beam, as 
proposed by \cite{Met03}. In the regime of small-scale events, where the WLFs are energized by low-energy ($<20$~keV) 
electrons, the electrons are unable to penetrate through the upper chromosphere. Therefore, for these events to 
demonstrate WL emission, a common belief is to allow for energy transportation from the upper chromosphere to 
the photosphere via the back-warming effect \citep[]{Mac89}.

\cite{Mat03} and \cite{Hud06} have shown, using space-based instrumentation, that WLFs are detectable in flare 
categories as low as C1.6. However, the mechanism describing the WL emission in such low-energy events is still 
inconclusive and open to debate \citep[]{Fle07}. Often, a strong temporal correlation between the impulsive 
stage of a flare and the resulting WL emission is observed \citep[]{Hud92}. Indeed, \cite{All05} demonstrate 
dramatic increases in optical continuum emission during the impulsive stage of simulated flares. However, the 
duration of these WL enhancements appear to be variable. \cite{Bop73} and \cite{Haw91}, through similar investigations of 
WLFs on dMe stars, have found very different continuum emission behaviour. The former find the WL 
enhancement ends once the impulsive stage of the flare has ceased, while the latter detect a gradual decay 
in continuum emission, even after the end of the impulsive phase.

Atmospheric seeing, which is an intrinsic part of all data sets 
acquired using ground-based telescopes, has previously prevented small-scale WL brightenings from being 
detected \citep[]{Hie82}. However, due to the high-order adaptive optics (AO) employed on large telescopes in 
conjunction with proven image reconstruction 
algorithms, small-scale structures can now be observed with unprecedented spatial and temporal resolution. 
In this letter, we utilize a high-cadence data set to search for flare-induced WL emission originating in the 
solar photosphere from a very modest C2.0 flare.

\section{Observations}
\label{observations}

\begin{figure*}
\epsscale{0.85}
\plotone{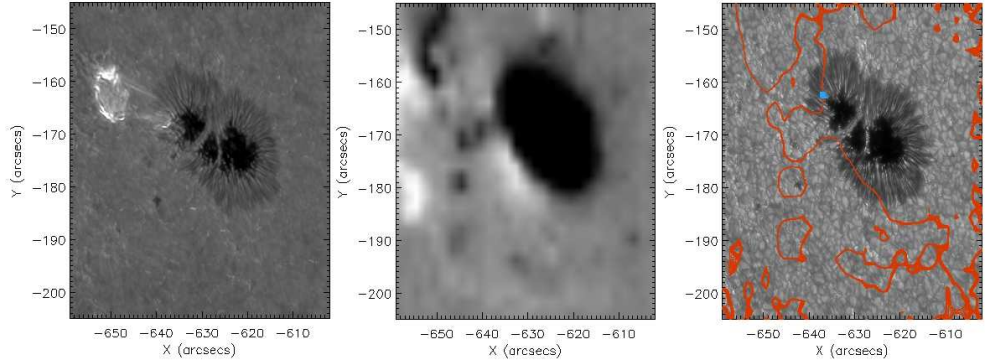}
\caption{Simultaneous images taken in the red continuum (left) and blue continuum (right) during a C2.0 
flare event (07:51:39~UT). The middle panel is an {\sc{mdi}} magnetogram taken at the closest time to the flare 
(11:15:02~UT) which shows regions of rapidly changing magnetic field strength and polarity. The left panel 
shows the extent of H$\alpha$ core-emission contamination during the flare as a result 
of the filter sampling a large portion of the H$\alpha$ line profile. The blue continuum filter is 
far removed from any chromospheric profiles and is therefore an accurate representation of the white-light continuum. 
The right image, taken at the time of the flare, details the location of white-light emission 
(blue contour) as well as a trace of the apparent polarity inversion lines (red contours) derived from {\sc{mdi}} data 
close to the time of the flare. The scale is in heliocentric coordinates where $1\arcsec \approx 725$~km.
\label{f1}}
\end{figure*}

The data presented here are part of an observing sequence obtained on 2007 August 24, with the Swedish Solar Telescope (SST) 
on the island of La~Palma. The optical setup allowed us to image a $68\arcsec \times 68\arcsec$ region surrounding active 
region NOAA~10969 complete with solar rotation tracking. This active region was located at heliocentric 
co-ordinates ($-516\arcsec$,$-179\arcsec$), or S05E33 in the solar NS-EW co-ordinate system. The Solar Optical Universal 
Polarimeter (SOUP) was implemented to provide 2-dimensional spectral information across the H$\alpha$ line 
profile centred at $6562.8$~{\AA}. In addition, a series of Ca~{\sc{ii}} interference filters were used to provide 
high-cadence imaging in this portion of the optical spectrum. 
For the purpose of this letter, only red (H$\alpha$) continuum and blue (Ca~{\sc{ii}}) continuum observations will be presented.

The observations employed in the present analysis consist of 39000 images in each wavelength, taken with a  0.12~s cadence, 
providing just over one hour of uninterrupted data. The acquired H$\alpha$ continuum images have a sampling of $0.068\arcsec$ per 
pixel, to match the telescope's diffraction-limited resolution to that of the $1024 \times 1024$~pixel$^{2}$ CCD. Images of 
the Ca~{\sc{ii}} continuum were captured using a $2048 \times 2048$~pixel$^{2}$ CCD with a sampling of $0.034\arcsec$ per 
pixel. Although the Ca~{\sc{ii}} camera was oversampled, this was deemed necessary to keep the dimensions of the 
field-of-view the same for both cameras. Eighth-order AO \cite[]{Sch03} were used throughout the data 
acquisition. The acquisition time for this observing sequence was early in the 
morning, and seeing levels were excellent with minimal variation throughout the time series. During the observing sequence a 
C2.0 flare was observed, originating from NOAA~10969 at 07:49~UT.  

Multi-Object Multi-Frame Blind Deconvolution \citep[MOMFBD;][]{van05} image restoration was implemented to 
remove small-scale atmospheric distortions present in the data. Sets of 80 exposures were included in the restorations, producing 
a new effective cadence of 9~s. All reconstructed images were subjected to a Fourier co-aligning routine, in which cross-correlation 
and squared mean absolute deviation techniques are utilized to provide sub-pixel co-alignment accuracy. In addition, full-disk 
{\sc{mdi}} magnetogrames are used to study changes in the line-of-sight magnetic field during the observations.

\section{Analysis and Results}
\label{analy}

Due to the width of the red continuum filter (10~{\AA} centered at 6565.2~{\AA}), it includes the full  
H$\alpha$ line profile. During the observed flare, the resulting images therefore include a significant contribution from 
the H$\alpha$ core. Therefore, these data can be compared with the blue continuum (10~{\AA} filtergram centered at 3953.7~{\AA}), 
whose images are not 
contaminated by any chromospheric emission. The left panel of Figure~\ref{f1} indicates the clear-cut 
nature of H$\alpha$ line-core contributions to the continuum filtergrams during the observed flare sequence. A 
simultaneous blue continuum image is displayed in the right panel of Figure~\ref{f1}, yet due to the 
bandpass of the chosen interference filter, no chromospheric contributions are visible.

\begin{figure}
\epsscale{1.0}
\plotone{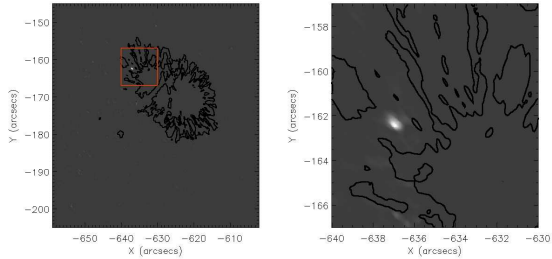}
\caption{Difference image formed by subtracting a quiescent blue continuum image (taken at 07:50:30) from one obtained 
during the impulsive stage of the observed white light flare (taken at 07:51:39). The outlines of the penumbra and small 
pore (just south of the sunspot) are contoured in grey in the left panel. A strong brightening is visible in the 
north-eastern edge of the penumbra, caused by white light emission during the flare. An expanded portion of the site 
of the flare kernel is shown in the right panel, as denoted by the red box.
\label{f2}}
\end{figure}

In order to pinpoint any brightenings in the blue continuum data set as a result of the flare, difference imaging 
techniques are implemented \citep[see e.g.][]{Asc99}. By subtracting a quiescent continuum image from one undergoing 
a flare brightening, small-scale intensity enhancements become more obvious. Figure~\ref{f2} shows the brightening present 
in the blue continuum data set at the time of the flare. When compared to the footpoint of the H$\alpha$ flare ribbon shown in 
Figure~\ref{f1}, it is clear that the blue continuum brightening originates from the same location in the sunspot 
penumbra. Indeed, the location of the flare kernel is consistent with the results of a statistical study by \cite{Mat03}, 
who find a number of small-flare kernels located near the penumbra/photosphere boundary.
The blue continuum brightening occupies 106~pixels in an approximately circular shape, based on a $5\sigma$ detection 
threshold above the quiescent mean. Assuming this circular geometry, a kernel diameter of $\approx 300$~km (12 pixels) 
exists covering a total area of $\approx 70000$~km$^{2}$. 
A flare kernel of these dimensions will therefore be undetectable by the space-based instrumentation used in the 
\cite{Mat03} and \cite{Hud06} studies. These surveys struggled to detect C-class WLF emission, primarily due to the 
limited spatial resolution offered by the {\sl{TRACE}} and {\sl{Yohkoh}} instruments, which is 725~km at best. 

\begin{figure}
\epsscale{1.0}
\plotone{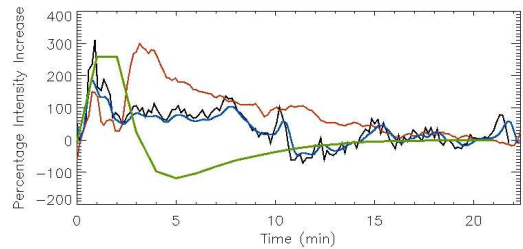}
\caption{Intensity plots of the flare kernel evolution in the red continuum (solid red line), MOMFBD-reconstructed blue 
continuum (solid black line), derivative of {\sl{GOES}} soft X-rays (solid green line) in arbitrary units and raw blue continuum 
(solid blue line) averaged over 9~s to maintain consistency with the reconstructed data and to remove small-scale 
fluctuations. Time from the start of the observing sequence (07:50:30~UT) is plotted 
along the x-axis, while the corresponding intensities, as a percentage of the quiescent flare site, are displayed along the 
y-axis. In each instance, the light curve is generated for the brightest flare-kernel pixels, i.e. those superceding a 
threshold equal to the quiescent background plus $8\sigma$. Note how the continuum brightening is directly correlated 
in all three intensity plots with the derivative of the {\sl{GOES}} soft X-ray flux, consistent with the Neupert effect. 
The large brightening in the H$\alpha$ plot is a result of chromospheric line-core emission contaminating the filtergram.
\label{f3}}
\end{figure}

Examining the light curve of the blue continuum brightening, plotted as a solid black line in Figure~\ref{f3}, it is 
clear that the intensity increase correlates directly with the heightened red continuum emission, plotted in the 
upper panel. The associated peak in the blue continuum is $\approx$~300\% above the  
quiescent value. This is significantly higher than the typical $5 - 50$\% increases detected in previous analyses of WLFs 
\citep[e.g.][]{Lin76}. We believe that this large increase is an effect of the high spatial resolution used in this study. 
By simply degrading the spatial resolution of the current data to that of {\sl{TRACE}} \citep[]{Han99}, we find that the peak 
blue continuum brightening is reduced to only $\approx 1$\% above its quiescent value. This small increase in luminosity renders 
the brightening below the values typically observed using satellite-based instruments, since a fluctuation of $\approx 1$\% 
will be lost amongst detector noise. It is also interesting to note, from Figure~\ref{f3}, that the associated peak in the raw 
blue continuum is only $\approx$~200\% above its quiescent value compared to the $\approx$~300\% increase found when investigating 
the MOMFBD-reconstructed data. This difference can be attributed to the restoration of diffraction-limited resolution 
after implementation of MOMFBD. An identical intensity threshold applied to reconstructed images will identify fewer pixels (due 
to an improved spatial resolution), but with a higher overall intensity when compared to pre-reconstructed data. For 
example, using a lower limit threshold of the quiescent mean plus $8\sigma$, MOMFBD-reconstructed data at the time of 
the flare isolates only 36~pixels corresponding to the brightest regions of the flare kernel, whereas pre-reconstructed 
continuum data extracts 57~pixels. For the 36~pixels isolated in 
the MOMFBD-reconstructed data, the average intensity is $\approx$~300\% above the quiescent mean, whereas an average 
increase of $\approx$~200\% is found for the 57~pixels extracted in the pre-reconstructed data set. This indicates the degree 
of intensity ``smearing'' caused by the Earth's turbulent atmosphere, but through proven reconstruction algorithms this can 
be removed to restore the image, and hence intensity, to its true diffraction-limited resolution.

The detected contrast enhancement can provide some insight into the origin of the WL 
continuum brightening. \cite{Din99} and \cite{Din03} carried out detailed non-LTE 
calculations on solar atmospheres bombarded by electron beams and heated 
by backwarming, with these calculations performed for several 
heliocentric distances. Their main finding is that the highest contrast is 
produced near the limb for a heated atmosphere that does not involve 
non-thermal effects from the electron beam. The actual value for the measured contrast 
is considerably higher than those predicted by the models and may seem well in excess of 
the values predicted by current radiative transfer calculations. However, we need to 
emphasize that since localized flare temperatures are high, it is easier for WL emission to be observed 
in the generally cooler atmosphere of a sunspot \citep[]{Din96}. The high contrast ($\approx~300$\%) quoted 
above has been estimated relative to quiescent penumbral intensities. If the contrast is determined with 
respect to typical granulation values, it is reduced to 24\%. We therefore conclude that the high contrast 
observed in this study remains consistent with current non-LTE calculations and supports heating through 
radiative back-warming models \cite[see also][]{Liu01}.

The initial impulsive stage of the flare is clearly visible in both the red and blue continuum plots. However, the red 
intensity plot also reveals a signature of H$\alpha$ line-core contamination. 
The large intensity spike after $\approx 4$~min is a result of H$\alpha$ core emission in the filtergrams 
(see also left panel of Fig.~\ref{f1}). As expected, this form of ``chromospheric contamination'' is absent in the blue continuum 
images. From Figure~\ref{f3}, it is clear that the true continuum brightening acts as a precursor to the influx of 
H$\alpha$ core emission, with an associated delay time of 2~min. This is consistent with the theoretical work of 
\cite{All05}, whose simulated flare events consist of impulsive WL brightenings followed by line-core emission, often 
with delays of $\approx 2$~min. Furthermore, the continuum brightening lasts less than 2~min in both the red and 
blue channels. The WLF emission is therefore short lived and only present during the impulsive stage of the 
flare. Through examination of an M-class flare, \cite{Che06} find WLF emission during the impulsive phase, with 
very little residual flux in the latter stages of the flare event. Therefore, the data presented here corroborate the 
work of \cite{Bop73}, providing a good example of impulsively-generated, short-lived continuum emission. In many flare cases, 
WL emission is compared directly to the derivative of soft X-ray flux. If the Neupert effect \citep[e.g.][]{Nin08} is valid, 
a direct correspondence between the derivative of the soft X-ray flux and pure WL emission should be evident. Utilizing soft 
X-ray data measured by the {\sl{GOES}} spacecraft, we determine that the Neupert effect is valid during the presented flare 
event (Fig~\ref{f3}). However, the cadence of the {\sl{GOES}} instrument is 1~min, so timing errors associated with soft 
X-ray investigations will be considerably larger than those based upon the continuum data presented here.

With the location of the flare kernel identified, it is important to investigate how the magnetic-field morphology ties in with the 
WL continuum brightening. Typically, flare kernels originate from locations of strong magnetic field gradient, 
where the field strength changes abruptly over a very short distance \citep[]{Moo85}. Additionally, it is of 
interest to examine where the magnetic energy is released as a result of the flare, and in what direction this release 
takes place. A two-dimensional magnetic field gradient can be constructed from {\sc{mdi}} data obtained just prior to the 
flare, when the magnitude of the gradient should be at its greatest \citep[]{Cui07}. Furthermore, by taking a difference 
image (pre-flare subtracted from post-flare) of {\sc{mdi}} magnetic field information, an accurate representation of changes to the 
line-of-sight magnetic field can be constructed. Following this form of analysis, we find the observed continuum flare kernel 
is positioned within a region of strong magnetic field gradient and that the magnetic energy release (from {\sc{mdi}} difference 
imaging) is directly correlated with the H$\alpha$ flare ribbon displayed in the left panel of Figure~\ref{f1}. This is consistent 
with recent X-class flare observations \citep[]{Guo08,Jin08}, although on a much smaller scale. 

From examination of the right panel of Figure~\ref{f1}, it appears that the detected flare kernel is located 
co-spatially with a polarity inversion line. \cite{Bel04} find that the inclination of the mean magnetic field near 
the penumbral boundary reaches $70 - 80^{\circ}$, implying that active regions observed 
away from disk center often show a false polarity inversion line. In such cases, the apparent neutral line 
falls within the so-called limb-side penumbra \citep[]{Sai05}, as observed in the right panel of Figure~\ref{f1}. 
Penumbral models involving two-component magnetic field inclinations produce two distinct neutral lines. Magnetic field 
vectors with more horizontal components will show an apparent neutral line lying close to the umbra, whereas a more 
vertical magnetic field component will display a polarity inversion line closer to the penumbral boundary. This dual 
neutral-line configuration is visible in the North-East part of Figure~\ref{f1}, close to where the WL brightening originates. 
The true neutral line will lie somewhere in between, with the precise location depending on the weightings made by each of 
the two components \citep[]{Tit93}. We therefore conclude that the observed WL brightening originates very close to the 
true polarity inversion line, often the case for large-scale WLFs \citep[]{Li05}.

As detailed above, the detected flare kernel has a diameter of $\approx~300$~km, thus 
placing this event out of the range of most current ground- and space-based observatories. As discussed 
in \cite{Mat03} and demonstrated here, there is no reason why WLF emission should not be produced in all flares. It is 
only the sensitivity and resolution of telescope instrumentation which limits the lower WL emission detection threshold. 
Therefore, to probe these events and conclusively derive WL emission characteristics, observers must utilize high spatial 
and temporal resolution data sets covering the wide energy range of solar flares.

\section{Concluding Remarks}
\label{conc}

The unprecedented spatial and temporal resolution data acquired with the SST has revealed WLF emission, lasting 
approximately 2~min, originating from a compact C2.0 flare. The localized flare emission peaks at 300\% above the 
quiescent flux, much larger than previously observed in satellite-based data sets, simply as a consequence of the high 
spatial resolution achieved here. Blue continuum images clearly show a bright flare kernel located close to a region of 
magnetic polarity inversion. The WLF brightening is present during the impulsive stage 
of the flare, with the effects of core emission subsequently delayed by $\approx~2$~min. This corroborates 
current theoretical models \citep[e.g.][]{All05} related to the generation of optical flare emission.
Furthermore, the observed H$\alpha$ flare ribbon correlates spatially with the 
location of rapidly changing magnetic energy. Additionally, the footpoint of the H$\alpha$ flare ribbon is co-spatially 
located with the blue continuum brightening, thus establishing the observed WL emission as a flare-induced 
phenomena, consistent with radiative back-warming models.

\acknowledgments

DBJ is supported by a Northern Ireland Department for Employment and Learning studentship. DBJ additionally thanks NASA 
Goddard Space Flight Center for a CAST studentship. FPK is grateful to AWE Aldermaston for the award of a William Penney 
Fellowship. The Swedish 1-m Solar Telescope is operated on the island of La Palma by the Institute for Solar Physics of 
the Royal Swedish Academy of Sciences in the Spanish Observatorio del Roque de los Muchachos of the Instituto de Astrof\'{i}sica 
de Canarias. These observations have been funded by the Optical Infrared Coordination network (OPTICON), a major international 
collaboration supported by the Research Infrastructures Programme of the European Commissions Sixth Framework Programme.

\end{document}